\documentclass[journal]{IEEEtran}
\usepackage[utf8]{inputenc} 
\usepackage{mathtools}
\usepackage{authblk}

\usepackage{cite}
   \usepackage[pdftex]{graphicx}
   \DeclareGraphicsExtensions{.pdf,.jpeg,.png}

\usepackage{program}
\usepackage{algorithm} 
\usepackage{algpseudocode}

\usepackage{balance}
\usepackage{xcolor}
\algnewcommand{\IIf}[1]{\State\algorithmicif\ #1\ \algorithmicthen}
\algnewcommand{\EndIIf}{\unskip\ \algorithmicend\ \algorithmicif}
\algnewcommand{\IFor}[1]{\State\algorithmicfor\ #1\ \algorithmicthen}
\algnewcommand{\EndIFor}{\unskip\ \algorithmicend\ \algorithmicfor}

\begin{document}

\pagenumbering{gobble}

\title{Implementation of binary stochastic STDP learning using chalcogenide-based memristive devices}


\author{C. Mohan}
\author{L. A. Camuñas-Mesa}
\author{J. M. de la Rosa}
\author{T. Serrano-Gotarredona}
\author{B. Linares-Barranco}

\affil{Instituto de Microelectrónica de Sevilla (IMSE-CNM), CSIC y Universidad de Sevilla, Sevilla, Spain}

\renewcommand\Authands{ and }

\maketitle

\begin{abstract}
The emergence of nano-scale memristive devices encouraged many different research areas to exploit their use in multiple applications. One of the proposed applications was to implement synaptic connections in bio-inspired neuromorphic systems. Large-scale neuromorphic hardware platforms are being developed with increasing number of neurons and synapses, having a critical bottleneck in the online learning capabilities. Spike-timing-dependent plasticity (STDP) is a widely used learning mechanism inspired by biology which updates the synaptic weight as a function of the temporal correlation between pre- and post-synaptic spikes. In this work, we demonstrate experimentally that binary stochastic STDP learning can be obtained from a memristor when the appropriate pulses are applied at both sides of the device.  

\end{abstract}


\IEEEpeerreviewmaketitle

\section{Introduction}
\label{sec:intro}


\IEEEPARstart{N}{euromorphic} engineering has demonstrated in recent decades a tremendous potentiality to develop bio-inspired computing systems for high-speed processing with low-power consumption \cite{Indiveri}. Different kinds of neural networks have been proposed, with Spiking Neural Networks (SNNs) being specially suited for low-power hardware implementation, given that neurons are active only when input spikes are received, which improves computational efficiency \cite{Bouvier}. These networks mimic the main properties of the biological brain, like the massive parallel connectivity between layers of neurons. However, as the number of implemented neurons and synapses increases, the implementation of efficient learning algorithms becomes one of the critical bottlenecks in terms of resources, time and power consumption.

When memristive elements were first proposed theoretically \cite{Chua1} and later demonstrated experimentally \cite{HP}, they opened a promising new strategy for emulating biological synapses, as the well-known learning mechanism Spike-timing-dependent plasticity (STDP) \cite{Gerstner1} could be obtained naturally when applying pre- and post-synaptic spikes at both terminals of a memristor \cite{Linares,Zamarreno}. In order to reproduce different possible STDP rules, it is necessary to design spiking neurons which can generate programmable spike shapes \cite{ICECS,LuisAICAS}. 

In this work, we have demonstrated experimentally an STDP rule on a silver-chalcogenide based memristor (commonly referred to as Neuro-Bit) \cite{neuro} using the ArC ONE characterization platform \cite{arc}. To this end, we propose a combination of multi-spike pulse shapes for pre- and post-synaptic signals, taking advantage of the stochastic behavior of the device obtaining a learning probability curve that resembles a stochastic binary STDP rule (a specific kind of STDP rule where 1-bit weights change following a given probability \cite{Amir}). This can be a very promising strategy to implement online learning on hybrid memristor-CMOS systems \cite{Camunas}, providing learning capabilities with minimum nano-scale resources consumption.

\section{Memristors}
\label{sec:memr}
\begin{figure}[!t]
\centering
\includegraphics[width=1.8in]{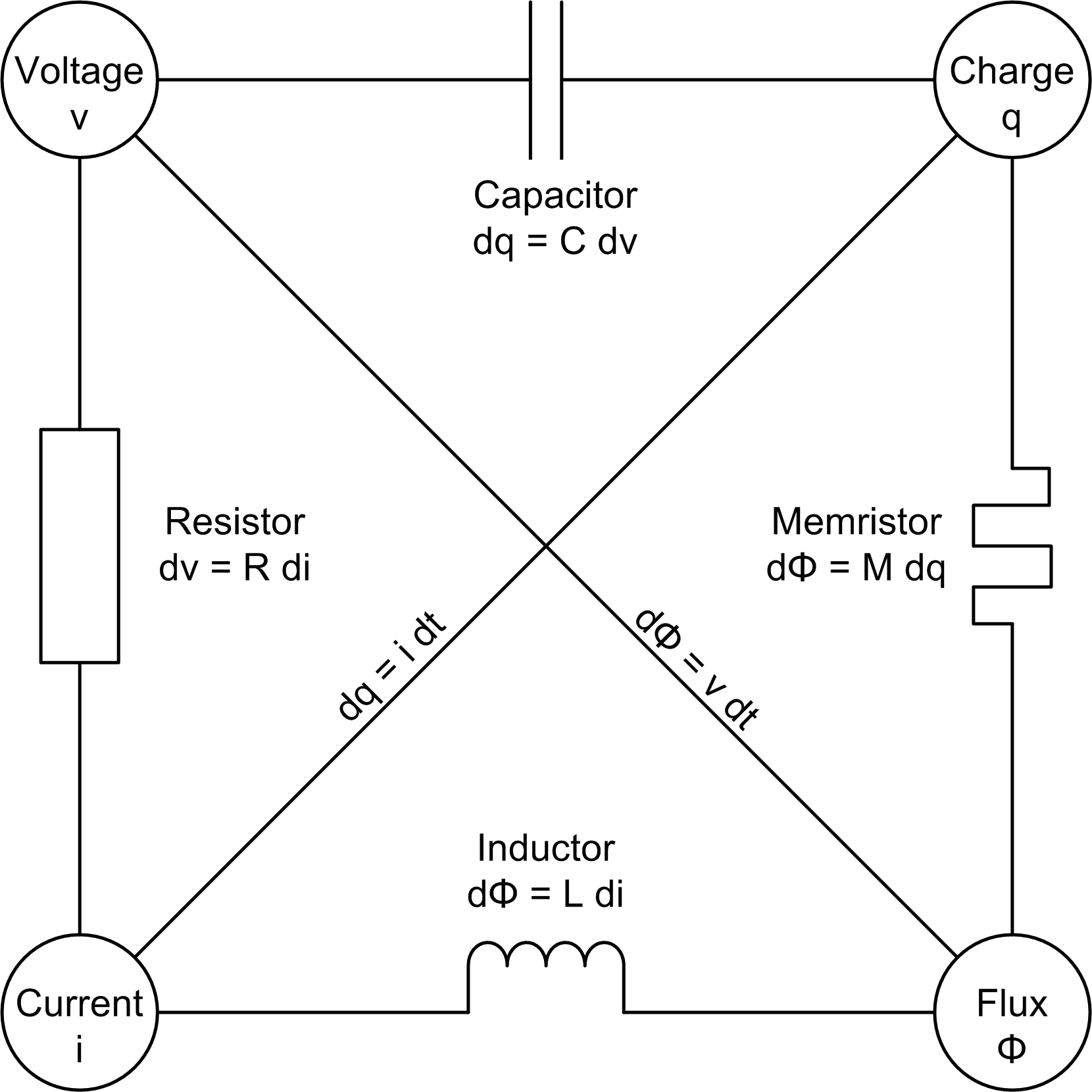}
\caption{Illustration of the four canonical elements in terms of the basic electrical quantities, with memristor on the right-hand side.}
\label{fig:Elements}
\end{figure}

The memristor is a non-linear passive two-terminal component postulated by Chua \cite{Chua1} in 1971. According to circuit theoretical fundamentals, there are four basic electrical quantities \cite{Chua2}: (1) voltage difference between two-terminals $v$, (2) current flowing into a device terminal $i$, (3) charge flowing through a device terminal or integral of current $q = \int i(\tau)d\tau$, and (4) flux or integral of voltage $\phi = \int v(\tau)d\tau$. A two-terminal device is said to be canonical \cite{Chua2} if either two of the four basic electrical quantities are related by a static relationship. Describing the traditional devices in terms of these electrical quantities, the resistor presents a static relationship between voltage $v$ and current $i$, the capacitor presents a static relationship between charge $q$ and voltage $v$, and the inductor presents a static relationship between current $i$ and flux $\phi$. Ignoring the combinations of a quantity with its own time derivative, leaves one more possibility: charge $q$ and flux $\phi$. This last combination led Chua to postulate the existence of the missing two-terminal element: the memristor, as illustrated in Fig. \ref{fig:Elements}. Following Chua's generalization of memristive systems \cite{Chua3}, the concept of memristor can be extended to any device exhibiting resistive behavior whose resistance can change through some of the four basic electrical quantities (or combinations of them), while exhibiting memory for that resistance. 

Decades after Chua's work, a team at HP Labs obtained a memristor experimentally for the first time, using a thin film of titanium oxide \cite{HP}, starting a new race to fabricate robust memristive devices and integrate them over CMOS platforms to implement hybrid systems  \cite{Camunas}.

\section{Spike-Timing-Dependent Plasticity}
\label{sec:stdp}

\begin{figure}[!t]
\centering
\includegraphics[width=2.5in]{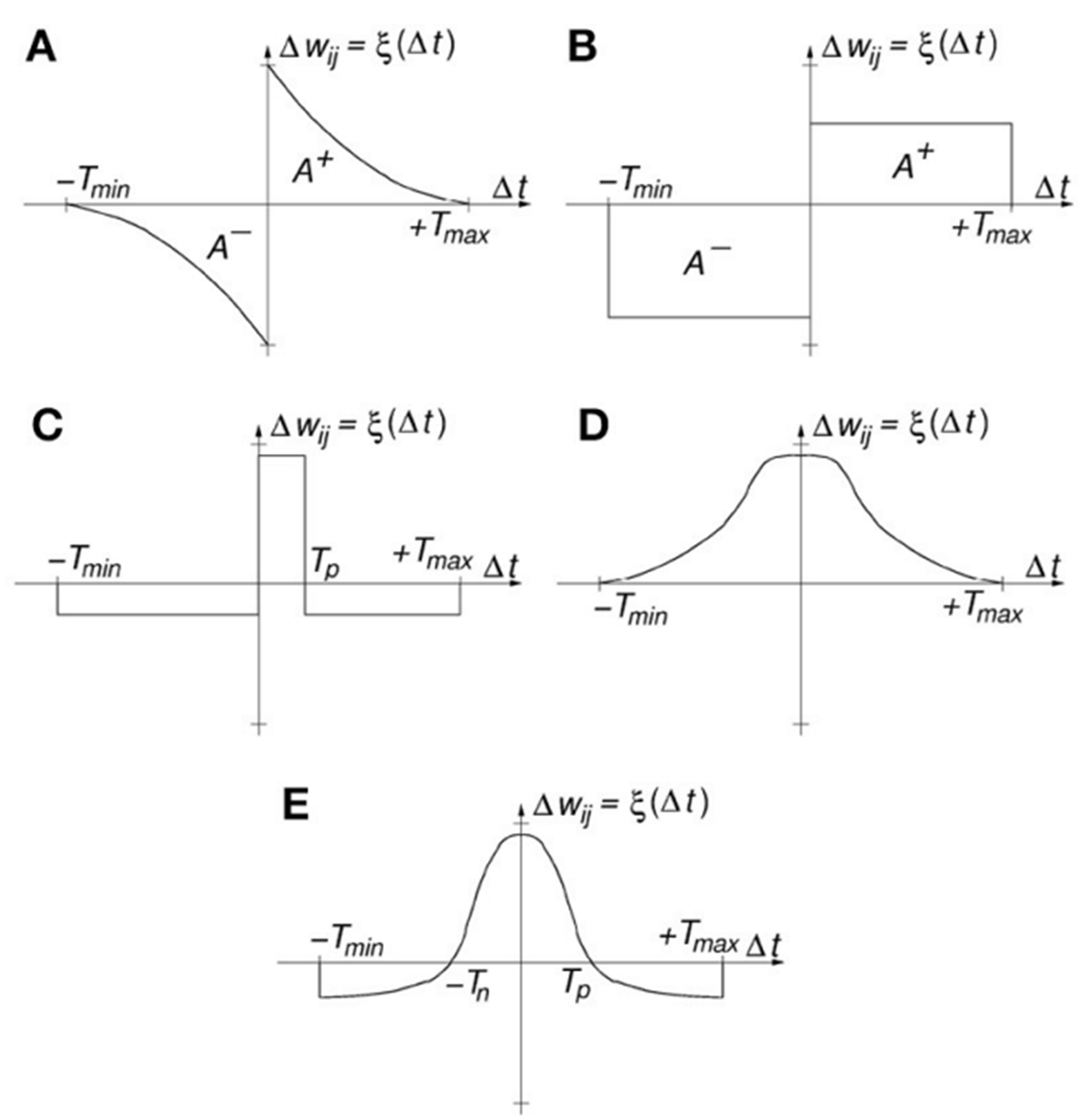}
\caption{Different time-based STDP functions.}
\label{fig:STDP}
\end{figure}

Spike-timing-dependent plasticity (STDP) learning algorithms were proposed in the 1990s \cite{Gerstner1,Gerstner2} as a particularization of the Hebbian learning rule formulated in the 1940s \cite{Hebb}. Considering a synapse with weight $w$ which connects two spiking neurons, the change in the synaptic weight $\Delta w$ is described as a function $\xi$ of the time difference $\Delta t$ between the post-synaptic spike at $t_{post}$ and the pre-synaptic spike at $t_{pre}$. Fig. \ref{fig:STDP} represents different STDP learning functions proposed for $\Delta w = \xi(\Delta t)$, with \textbf{A} showing the typical update function first measured in biology \cite{Poo} and \textbf{B}-\textbf{E} showing other STDP learning functions that have also been observed in biology or used by computational neuroscientists \cite{Bell}. In general, they all present some ranges of $\Delta t$ values where the function produces synaptic potentiation (positive $\Delta w$) and some other ranges of $\Delta t$ where it produces synaptic depression (negative $\Delta w$). Some functions present fixed potentiation or depression values for $\Delta w$, while some others follow certain curves. In general, they are all oriented towards obtaining continuous values of the synaptic weight $w$. 
 
\subsection{Binary stochastic STDP}
\label{subsec:binary}

Although STDP typically performs very small weight changes on synapses implying high-resolution implementations, several kinds of limitations (like hardware resources, memory or power consumption) recommend the reduction of weight resolution down to 1 single bit  and including some kind of stochastic weight update \cite{Amir}. That means that only ‘1’ or ‘0’ values are valid for each synaptic weight, with the algorithm changing a weight from ‘1’ to ‘0’ or vice versa following a probability given by the STDP rule \cite{Maas,Suri}. This strategy is especially interesting when implementing synapses with memristors, given the current difficulties to control precisely the analog resistive value of these devices. Memristors are typically used in binary mode for in-situ computing to overcome sharp switching threshold, functional uniformity, intermediate state initialization and state decay \cite{Borghetti,Pfeil,Waser}. This binary mode involves switching memristors between two possible resistive states, i.e., HRS (High Resistive State) and LRS (Low Resistive State). Therefore, we propose a binary stochastic STDP algorithm to be applied to memristive synapses. For that, some pre- and post-synaptic pulse shapes are defined in Section \ref{subsec:stdp_test}.

\section{Experimental results}
\label{sec:results}

\subsection{Experimental setup}
\label{subsec:setup}
\begin{figure}[!t]
\centering
\includegraphics[width=2.5in]{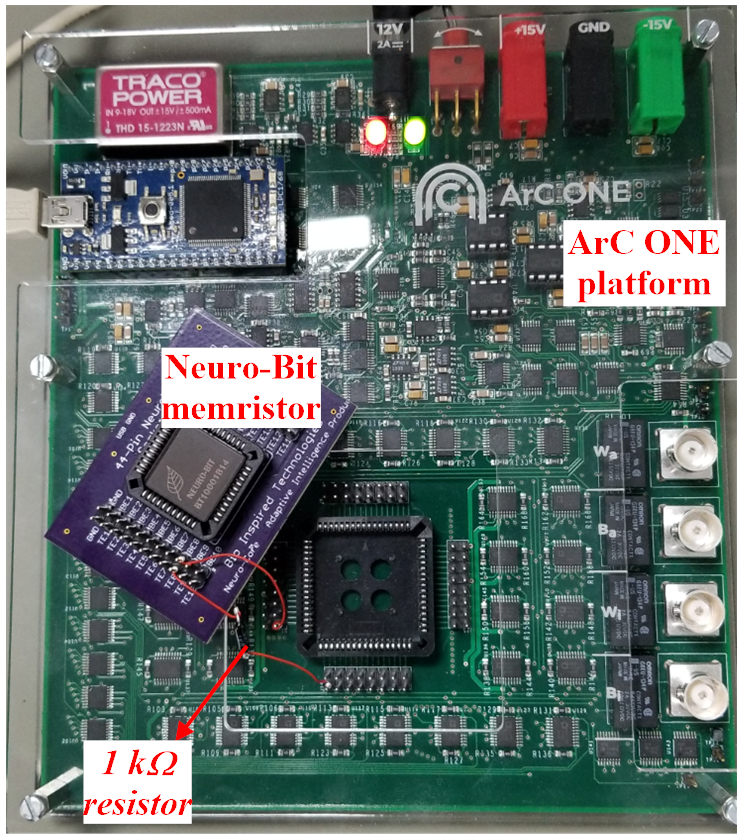}
\caption{A Neuro-Bit memristor connected to the active word-line and bit-line terminals of ArC ONE platform.}
\label{fig:arc}
\end{figure}

The experimental setup mainly comprises the ArC ONE platform and Neuro-Bit memristor, as shown in Fig.~\ref{fig:arc}. ArC ONE platform is used to characterize a single or an array of selector-less memristor devices either directly on wafers or in packed samples \cite{arc,berd,serb}. The ArC ONE board has an mBED microcontroller driver, bias generating opamps, sense resistor bank, read/write feedback buffers, programmable current source for current pulsing, TIA read opamps, PLCC68 DIP socket, header pins, power management block, resistor banks and digital components like decoders, multiplexers and switches. The working principle of ArC ONE platform is to pick the active and default wordlines and bitlines and apply DC voltage pulses for the needed operation. The user can perform operations such as form, write, erase and read in a sequence or a closed-loop to plot I-V characteristics, to do endurance and retention tests.

Neuro-Bit, a silver-chalcogenide based two-terminal device has a layer of Ge$_2$Se$_3$ (30 nm), Ag$_2$Se (50 nm), and Ag (50 nm)- sandwiched between the top and bottom electrodes \cite{hiro,wang,camp,camp1,camp2,oble}. The 44-pin PLCC breakout board of Neuro-Bit has 20 bonded memristor devices \cite{neuro}. The pre-programming resistance of the device is 50 M$\Omega$. A positive voltage sweep from 0 to 1 V with compliance current values between 100 nA and 30 $\mu$A can cause the device to switch to LRS (also called as `write' and typical LRS value = 8 k$\Omega$) at a certain voltage (called V$_{set}$) by making the Ag$^{+}$ ions to migrate into the active chalcogenide, Ge$_2$Se$_3$ layer and create a low resistance path through the insulator. Similarly, a negative voltage sweep from 0 to -1 V with compliance current values between 100 $\mu$A and 10 mA can cause the device to switch to HRS (also called as `erase' and typical HRS value = 13 M$\Omega$) at a certain voltage (called V$_{reset}$) by removing Ag$^{+}$ ions back from the active chalcogenide layer. The device is read by applying a read voltage, V$_{read}$ = 50 mV.

\subsection{Device Characterization}
\label{subsec:device}
The main challenges in using Neuro-Bit devices are the intrinsic variability and sensitivity of the device. Different samples of the device switch at different voltages V$_{set}$ and V$_{reset}$, and with different compliance current. Besides, for a single device, different behavior might be observed in two or more cycles of `write' or `erase'. These variations attribute to the stochasticity of the synaptic device, which results in the escape of local minima during learning and inference \cite{ackl}. Although the stochastic synapses raise reliability concerns in ANN, there are many embracing approaches such as- increasing the effective resolution of synaptic weights \cite{truenorth,gold,corr}, using stochastic neurons \cite{wije,shed}, etc. Neuro-Bit devices are also very sensitive. They can easily be permanently damaged when receiving voltage transients of few mV (including environmental noise), shorting the device or hanging it at LRS, when they are carelessly biased above the needed V$_{set}$, below the appropriate V$_{reset}$ or without the required compliance current. Hence, it is always recommended to carefully set-up the experimental test-bench and to attempt `write' and `erase' operations with very conservative values for each Neuro-Bit memristor until the user is comfortable with its performance.

\begin{figure}[!t]
\centering
\includegraphics[width=3.6in]{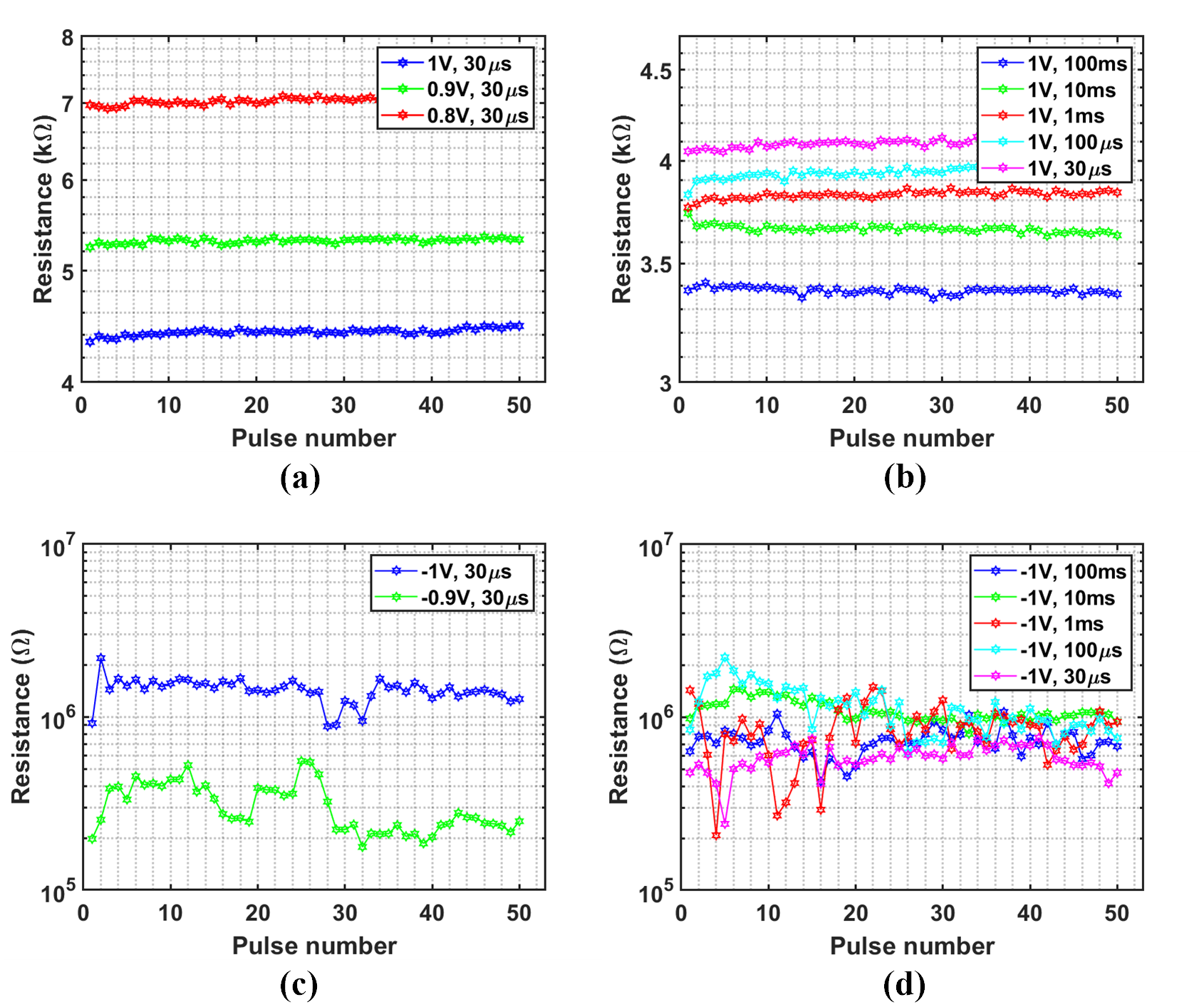}
\caption{Characterization of the device, showing the measured resistance after several Write and Erase pulses. (a) With the device initially in HRS, Write pulses are applied with a fixed width of $30 \mu s$ and different amplitudes. (b) With the device initially in HRS, Write pulses are applied with a fixed amplitude of $1V$ and different widths. (c) With the device initially in LRS, Erase pulses are applied with a fixed width of $30 \mu s$ and different amplitudes. (d) With the device initially in LRS, Erase pulses are applied with a fixed amplitude of $-1V$ and different widths.}
\label{fig:Write_erase}
\end{figure}

In order to characterize the Neuro-Bit memristor, the device is initially verified for I-V characteristic using HP4145 SPA, whose observation of DC voltage sweep measurements are documented in the user manual \cite{neuro}. After observing some successful repetitive switching in a Neuro-Bit device in HP4145 SPA, we connected it to the active wordline and bitline of ArC ONE platform through a 1 k$\Omega$ resistor (as a safety precaution to limit current during LRS) to characterize its switching using DC voltage pulses, as shown in Fig. \ref{fig:arc}. We varied both amplitude and pulse-width of the `write' and `erase' DC pulses using the ArC ONE Control GUI interface. Fig. \ref{fig:Write_erase} shows the LRS and HRS results for 50 `write' and `erase' pulses for different pulse-widths and amplitudes. The measured value for LRS is in the range of 3 to 7 k$\Omega$, which is quite similar to the typical value of 8 k$\Omega$ provided in the user manual. A high variability at HRS is observed, with values between hundreds of k$\Omega$ and 2 M$\Omega$, smaller than the typical value of 13 M$\Omega$. This difference is given by the limited precision of the measurements for such large resistance values.

\subsection{STDP tests}
\label{subsec:stdp_test}
\begin{figure}[!t]
\centering
\includegraphics[width=3.5in]{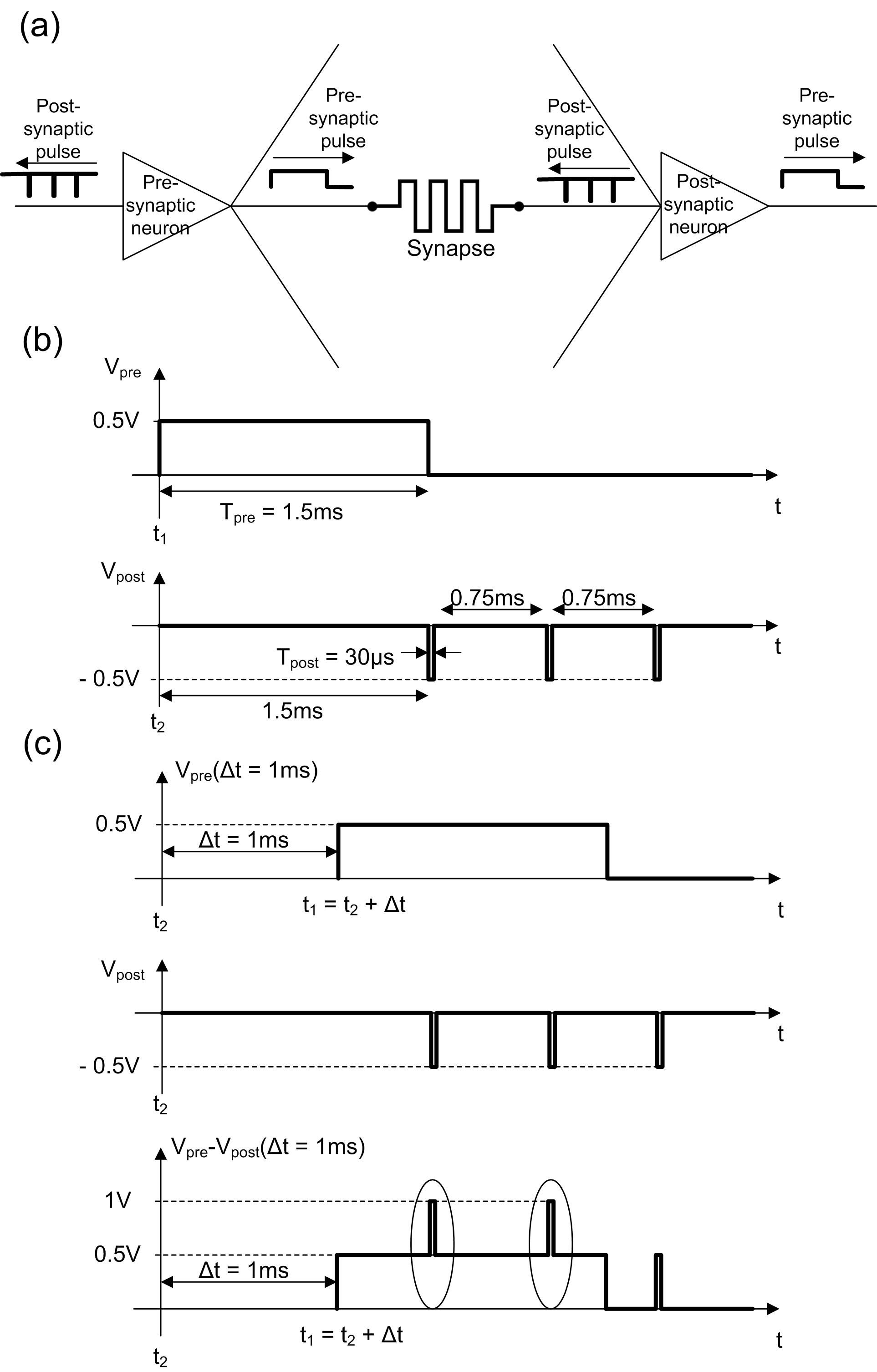}
\caption{(a) Illustration of the STDP rule applied to a synapse connecting two neurons. (b) Proposed pulse shapes for pre- and post-synaptic pulses. (c) Calculation of the applied pulse to the memristor ($V_{pre}-V_{post}$) with a time shift of $1ms$.}
\label{fig:Pre_Post}
\end{figure}

The main goal of this work is to demonstrate experimentally an STDP rule on Neuro-Bit memristor using the ArC ONE platform. As illustrated in Fig. \ref{fig:Pre_Post}(a), we use a single memristor to model a synapse connecting a pre- and a post-synaptic neuron. Each neuron generates pre-synaptic pulses $V_{pre}$ which propagate forward and post-synaptic pulses $V_{post}$ which propagate backwards. Therefore, the voltage across the synapse will be given by the difference $V_{pre}-V_{post}$. As described in Section \ref{sec:stdp}, an STDP rule updates the synaptic weight according to the time difference $\Delta t$ between $V_{pre}$ and $V_{post}$. In this work, we propose the pulse shapes shown in Fig. \ref{fig:Pre_Post}(b), where $V_{pre}$ is a positive pulse with amplitude $0.5V$ and width $1.5ms$, and $V_{post}$ is a train of $3$ negative pulses with amplitude $-0.5V$, width $30 \mu s$ and separation $0.75ms$. While the positive pulse in $V_{pre}$ starts at the onset time $t_1$, the first negative pulse in $V_{post}$ is delayed $1.5ms$ since the onset time $t_2$. With these particular shapes, we guarantee that the voltage applied to the memristor $V_{pre}-V_{post}$ will only reach $1V$ for a certain range of time difference $\Delta t$. As we observed during the characterization of the device, a pulse with $1V$ amplitude and $30 \mu s$ width is enough to write the memristor, i. e., to set its resistance to LRS, while pulses with $0.5V$ amplitude should not affect its state. An example is shown in Fig. \ref{fig:Pre_Post}(c), where a time difference of $\Delta t = 1ms$ is considered. In this case, a pre-synaptic pulse is generated at time $t_1 = t_2 + \Delta t$, while a post-synaptic pulse is generated at time $t_2$. With this particular time difference, the resultant voltage applied to the memristor is shown below, with two highlighted positive pulses with amplitude $1V$ and width $30 \mu s$, which should write the state of the device. In this case, the third $30 \mu s$ pulse should not affect the device.

\begin{figure}[!t]
\centering
\includegraphics[width=3.5in]{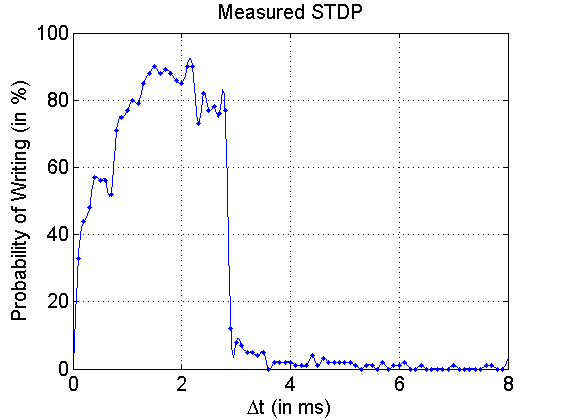}
\caption{Probability of `writing' the memristor for different values of $\Delta t$ between pre- and post-synaptic spikes, after testing the same stimulus $100$ times.}
\label{fig:PWriting}
\end{figure}

In order to reproduce the STDP rule, it was necessary to apply combinations of $V_{pre}$ and $V_{post}$ pulses to the Neuro-Bit memristor while sweeping the time difference between them $\Delta t$ using the ArC ONE platform. For the proposed pulses, a sweeping range from $0$ to $8ms$ was considered, with $0.1ms$ steps. For each value of $\Delta t$, the memristor was initially reset to HRS, then applied the pre- and post-synaptic pulses, and afterwards the resistance was read and stored. If the measured resistance was lower than 50 k$\Omega$, we considered the device was at LRS; otherwise, we assumed HRS. We repeated the whole process $100$ times, and we calculated the probability of writing for each value of $\Delta t$. The obtained result is shown in Fig. \ref{fig:PWriting}. If the behavior of the memristor were deterministic, we would expect to obtain $100\%$ probability for the range $0 < \Delta t < 3ms$ (shifting $V_{post}$ up to 3ms provides coincidence with at least one of the narrow pulses in $V_{pre}$), and $0\%$ otherwise. However, the stochasticity inherent to the device produced the STDP rule shown in the figure, with a maximum value of around $90\%$ around the middle of this range. For a certain range around $\Delta t = 1.5ms$, there are at least $2$ pulses with $1V$, therefore increasing the probability of writing. For values of $\Delta t$ outside the desired range, we observe a probability slightly larger than $0$ (around $1$ or $2\%$), representing the chance that the device can be written with applied voltages as low as $0.5V$.

In this example, we have obtained experimentally a binary stochastic STDP rule by using some specific pulse shapes, whose parameters can be adapted to modify the desired rule. Different STDP rules can be obtained by choosing the appropriate pulse shapes. 

\section{Conclusion}
\label{sec:conclusion}
A silver-chalcogenide based memristor (Neuro-Bit) has been characterized using ArC ONE platform, identifying the minimum pulse width and amplitude needed to write and erase the state of the device. Using this knowledge, an STDP learning rule has been measured experimentally by applying the appropriate pulses to both terminals of memristors, emulating the pre- and post-synaptic spikes in a biological system. This result demonstrates that different STDP rules can be implemented in hybrid memristor-CMOS systems, which can be crucial to perform large-scale online learning with minimum nano-scale resources consumption. 

\section*{Acknowledgment}

This work was funded by ICON (G0086), EU H2020 grants 824164 “HERMES”, 871371 “Memscales”, 871501 “NeurONN”, PCI2019-111826-2 “APROVIS3D”, 899559 “SpinAge” and by Spanish grants from the Ministry of Science and Innovation PID2019-105556GB-C31 (NANOMIND) and PID2019-103876RB-I00 (CORDION), Ministry of Economy and Competitivity TEC2015- 63884-C2-1-P (COGNET) (with support from the European Regional Development Fund), and Junta de Andalucía US-1260118 (Neuro-Radio). L. A. Camuñas-Mesa was funded by the VI PPIT through the Universidad de Sevilla.

\ifCLASSOPTIONcaptionsoff
  \newpage
\fi

\balance


\begin{thebibliography}{1}

\bibitem{Indiveri}
G. Indiveri et al., “Neuromorphic silicon neuron circuits,” \textit{Frontiers in
neuroscience}, 5 (73), 2011.

\bibitem{Bouvier}
M. Bouvier, A. Valentian, T. Mesquida, F. Rummens, M. Reybox, E. Vianello, E. Biegne,  “Spiking neural networks hardware implementations and challenges: a survey,” \textit{ACM Jorunal on Emerging Technlologies in Computing Systems}, 15 (2), 1–35, 2019.

\bibitem{Chua1}
L. O. Chua, “Memristor- The missing circuit element,” \textit{Transactions on Circuits theory IEEE}, vol. CT-18, no. 5, pp. 507-509, Sept. 1971.

\bibitem{HP}
D. B. Strukov, G. S. Snider, D. R. Stewart, R. S. Williams, R.S. “The missing memristor found,” \textit{Nature}, 453, 80–83, 2008.

\bibitem{Gerstner1}
W. Gerstner, R. Ritz, J. L. Hemmen. “Why spikes? Hebbian learning and retrieval of time-resolved excitation patterns,” \textit{Biol. Cybern.}, 69, 503–515, 1993.

\bibitem{Linares}
B. Linares-Barranco, T. and Serrano-Gotarredona, T., “Memristance can
explain spike-time-dependent-plasticity in neural synapses,” \textit{Nature precedings}, 2009.

\bibitem{Zamarreno}
C. Zamarreno-Ramos et al., “On spike-timing-dependent-plasticity, memristive devices, and building a self-learning visual cortex,” \textit{Frontiers in Neuroscience}, 5 (26), 2011.

\bibitem{ICECS}
T. Serrano-Gotarredona, B. Linares-Barranco, B., “Design of adaptive
nano/CMOS neural architectures,” \textit{19th IEEE International Conference on Electronics, Circuits, and Systems (ICECS)}, Seville, pp. 949-952, 2012.

\bibitem{LuisAICAS}
L. A. Camuñas-Mesa, B. Linares-Barranco and T. Serrano-Gotarredona, "Implementation of a tunable spiking neuron for STDP with memristors in FDSOI 28nm," \textit{2nd IEEE International Conference on Artificial Intelligence Circuits and Systems (AICAS)}, Genova, Italy, pp. 94-98, 2020.

\bibitem{neuro}
“Neuro-Bit memristor- user manual.” https://wiki.telavivmakers.org/images/0/0f/The\_Neuro-Bit\_Memristor\_user\_manual.pdf.

\bibitem{arc}
“ArC ONE Memristor Characterisation Platform.” http://www.arc-instruments.co.uk/products/arc-one/

\bibitem{Amir}
A. Yousefzadeh, E. Stromatias, M. Soto, T. Serrano-Gotarredona, B. Linares-Barranco,  “On Practical Issues for Stochastic STDP Hardware With 1-bit Synaptic
Weights,” \textit{Front. Neurosci.} 12:665, 2018.

\bibitem{Camunas}
L. A. Camuñas-Mesa, B. Linares-Barranco, T. Serrano-Gotarredona. “Neuromorphic Spiking Neural Networks and Their Memristor-CMOS Hardware Implementations,” \textit{Materials}, 12, 2745, 2019.

\bibitem{Chua2}
L. O. Chua, C. A. Desoer, and E. S. Kuh, “Linear and Nonlinear Circuits,” New York: McGraw-Hill. 1987.

\bibitem{Chua3}
L. O. Chua and S. Kang, “Memristive devices and systems,” \textit{Proceedings of the IEEE}, vol. 64, no. 2, pp. 209-223, 1976.

\bibitem{Gerstner2}
W. Gerstner, R. Kempter, J. Leo van Hemmen, H. Wagner. “A neuronal learning rule for sub-millisecond temporal coding,” \textit{Lett. Nat.}, 383, 76–78, 1996.

\bibitem{Hebb}
D. Hebb. “The Organization of Behavior,” Wiley: New York, NY, USA, 1949.

\bibitem{Poo}
G. Bi, M. Poo. “Synaptic modifications in cultured hippocampal neurons: dependence on spike timing, synaptic strength, and postsynaptic
cell type,” \textit{J. Neurosci.}, 18(24), 10464-10472, 1998.

\bibitem{Bell}
P. D. Roberts, C. C. Bell. “Spike timing dependent synaptic plasticity in biological systems,” \textit{Biol. Cybern.} 87, 392–403. 2002.

\bibitem{Maas}
 W. Maass, A. M. Zador, “Dynamic Stochastic Synapses as Computational Units”, \textit{Neural Comput.}, 11(4), 903-917, 1999.

\bibitem{Suri}
M. Suri, D. Querlioz, O. Bichler, G. Palma, E. Vianello, D. Vuillaume, et al., “Bio-inspired stochastic computing using binary cbram synapses,” \textit{IEEE Trans. Electron Devices} 60, 2402–2409, 2013.

\bibitem{Borghetti}
J. Borghetti, G. S. Snider, P. J. Kuekes, J. J. Yang, D. R. Stewart, R. S. Williams, “'Memristive' switches enable 'stateful' logic operations via material implications,” \textit{Nature} 464(7290), 873-876, 2010.

\bibitem{Pfeil}
T. Pfeil, T. C. Potjans, S. Schrader, W. Potjans, J. Schemmel, M. Diesmann, K. Meier, “Is a 4-bit synaptic weight resolution enough? - constraints on enabling spike-timing dependent plasticity in neuromorphic hardware,” \textit{Front. Neurosci.} 6, 2012.

\bibitem{Waser}
R. Waser, “Nanoelectronics and Information Technology,” \textit{Wiley- VCH}, Weinheim, 2012.

\bibitem{berd}
R. Berdan, A. Serb, A. Khiat, A. Regoutz, C. Papavassiliou and T. Prodromakis, “A $\mu$-controller-based system for interfacing selectionless RRAM crossbar arrays,” \textit{IEEE Transactions on Electron Devices}, vol. 62, pp. 2190-2196, July 2015.

\bibitem{serb}
A. Serb, A. Khiat and T. Prodromakis, “An RRAM Biasing Parameter Optimizer,” \textit{IEEE Transactions on Electron Devices}, vol. 62, pp. 3685-3691, November 2015.

\bibitem{hiro}
Y. Hirose and H. Hirose, “Polarity-dependent memory switching and behavior of Ag dendrite in Ag-photodoped amorphous \textsc{A}s$_2$\textsc{S}$_{3}$ films,” \textit{J. Appl. Phys.}, vol. 47, pp. 2767-2772, August 2008.

\bibitem{wang}
F. Wang, W. Dunn, M. Jain, C. D. Leo and N. Vickers, “The effects of active layer thickness on programmable metallization cell based on Ag-Ge-S,” \textit{Solid-State Electronics}, vol. 61, pp. 31-37, July 2011.

\bibitem{camp}
K. A. Campbell and J. T. Moore, “Silver-selenide/chalcogenide glass stack for resistance variable memory,” \textit{U.S. Patent 7, 151, 273}, December 2006.

\bibitem{camp1}
K. A. Campbell and J. T. Moore, “Resistance variable memory device and method of fabrication,” \textit{U.S. Patent 7, 348, 209}, March 2008.

\bibitem{camp2}
K. A. Campbell, “Methods of forming a PCRAM devie incorporating a resistance-variable chalcogenide element,” \textit{U.S. Patent 7, 354, 793}, April 2008.

\bibitem{oble}
A. S. Oblea, A. Timilsina, D. Moore and K. R. Campbell, “Silver chalcogenide based memristor devices,” \textit{The 2010 International Joint Conference on Neural Networks (IJCNN)}, pp. 1-3, July 2011.

\bibitem{ackl}
D. H. Ackley, G. E. Hinton and T. J. Sejnowski, “A learning algorithm for boltzmann machines,” \textit{Cognitive Science},  vol. 9, no. 1, pp. 147-169, 1985.

\bibitem{truenorth}
P. A. Merolla, J. V. Arthur, R. Alvarez-Icaza, A. S. Cassidy, J. Sawada, F. Akopyan, B. L. Jackson, N. Imam, C. Guo, Y.Nakamura, B. Brezzo, I. Vo, S. K. Esser, R. Appuswamy, B. Taba, A. Amir, M. D. Flickner, W. P. Risk, R. Manuhar and D. S. Modha, “A million spiking-neuron integrated circuit with a scalable communication network and interface,” \textit{Science},  vol. 345, pp. 668-673, August 2014.

\bibitem{gold}
D. H. Goldberg, G. Cauwenberghs and A. G. Andreou, “Probabilistic synaptic weighting in a reconfigurable network of VLSI integrate-and-fire neurons,” \textit{Neural Netw.},  vol. 14, pp. 781-793, 2001.

\bibitem{corr}
F. Corradi, C. Eliasmith and G. Indiveri, “Mapping arbitary mathematical functions and dynamical systems to neuromorphic VLSI circuits for spike-based neural computation,” \textit{2014 IEEE International Symposium on Circuits and Systems (ISCAS)}, pp. 269-272, 2014.

\bibitem{wije}
P. Wijesinghe, A. Ankit, A. Sengupta and K. Roy, “An all-memristor deep spiking neural computing system: A step toward realizing the low-power stochastic brain,” \textit{IEEE Transactions on Emerging Topics in Computational Intelligence}, vol. 2, no. 5, pp. 345-358, 2018.

\bibitem{shed}
M. Al-Shedivat, R. Naous, G. Cauwenberghs and K. N. Salama, “Memristors Empower Spiking Neurons With Stochasticity,” \textit{IEEE Journal on emerging and selected topics in circuits and systems}, vol. 2, no. 5, pp. 242-253, 2015.

\end{thebibliography}
\end{document}